\documentclass[11pt]{amsart}
\usepackage{amsfonts,amsmath,amssymb}
\usepackage[all]{xy}
\usepackage{hyperref}
\begin{document}

\newcommand{\ba}{{\bf a}}
\newcommand{\bb}{{\bf b}}
\newcommand{\ab}{{\rm ab}}
\newcommand{\bpi}{{\boldsymbol{\pi}}}
\newcommand{\be}{{\bf e}}
\newcommand{\oh}{{\mathfrak o}}
\newcommand{\m}{{\mathfrak m}}
\newcommand{\jnf}{{\rm inf}}
\newcommand{\A}{{\mathbb A}}
\newcommand{\cA}{{\mathcal A}}
\newcommand{\C}{{\mathbb C}}
\newcommand{\F}{{\mathbb F}}
\newcommand{\G}{{\mathbb G}}
\newcommand{\bP}{{\mathbb P}}
\newcommand{\R}{{\mathbb R}}
\newcommand{\SU}{{\rm SU}}
\newcommand{\Gal}{{\rm Gal}}
\newcommand{\Q}{{\mathbb Q}}
\newcommand{\bQ}{{\overline{\mathbb Q}}}
\newcommand{\T}{{\mathbb T}}
\newcommand{\Sp}{{\rm Sp \;}}
\newcommand{\W}{{\sf W}}
\newcommand{\sK}{{\sf K}}
\newcommand{\Z}{{\mathbb Z}} 
\newcommand{\ve}{{\varepsilon}}
\newcommand{\Spec}{{\rm Spec \:}}
\newcommand{\Hom}{{\rm Hom}}
\newcommand{\Mod}{{\rm Mod}}
\newcommand{\pt}{{\rm pt}}

\newcommand{\sslash}{{/\!/}}
\newcommand{\Det}{{\rm Det}}
\newcommand{\li}{{\rm li}}

\parindent=0pt
\parskip=6pt

\newcommand{\ie}{\textit{ie}\,}
\newcommand{\eg}{\textit{eg}\,}
\newcommand{\se}{{\sf e}}
\newcommand{\cf}{{\textit{cf}\,}}

\newcommand{\La}{{{\rm Lie}_\T}}

\title{Circular symmetry-breaking and topological Noether currents}

\author[J Morava]{J Morava}

\address{Department of Mathematics, The Johns Hopkins University,
Baltimore, Maryland} 

\email{jmorava1@jhu.edu}

\begin{abstract}{We propose a toy model for the algebraic topology of bubbling as circular symmetry-breaking, in terms of Noether currents and cobordism of manifolds with circle actions free along boundaries. This leads to an interpretation of 
Planck's radiation law \cite{9}(\S 2.4),\cite{13} as expressing the loss of a Noether symmetry when a bubble
\[
 |[S^2 \sslash \T]| \to {\rm pt}
 \]
 (analogous to blowing up or down in projective geometry) collapses; \cf the von K\'arm\'an street of sparks left when a candle flame wisps out \cite{15} .} \end{abstract}
 
\maketitle 

{\bf 1} In these notes `a space $X$ with circle action' will usually mean the geometric realization $|X/\T| := |[X/\T]|$
of the topological transformation groupoid defined by an action of the circle group $\T$ on the space $X$. 
The collapse map
\[
[X/\T] \to [{\rm pt}/\T]
\]
thus defines a ring homomorphism
\[
H^*_\T({\rm pt}) \cong \Z [c] \to H^*_\T(X)
\]
and therefore a natural characteristic class in the equivariant cohomology \cite{3}(\S 8) of $[X/\T]$, which I suggest  calling, the (topological) Noether charge of the space (\ie topological groupoid) $[X/\T]$, \ie with circle action \begin{footnote}{If ($M,\omega) \ni X)$ is a symplectic manifold w.c.a. there is a theory of moment maps $M \to \La$ \dots}\end{footnote}. The universal example $E\T$ then has charge $c \in H^2(B\T)$.

Since Gauss, the Chern class $c$ is understood as a measure of curvature. On the other hand, the integral Federer-de Rham current $b \in H_2(B\T,\Z)$ is (using Quillen's conventions) the class $[s^{-1}(0)]$ of the zero-locus of the canonical line bundle over $\C P^\infty$; under cap product it represents a Lefschetz hyperplane intersection operator, a $\delta$-function supported by the codimension two cycle $\C P^{n-1} \subset \C P^n, {\;} n \to \infty$. 

Kronecker duality pairs these by integration, but both terms make sense as elements of the (torsion-free) Swan - Tate cohomology $t_\T H\Z \cong \Z[c^{\pm 1}]$, where they can be compared over $\Q$. \bigskip

{\bf 2} The Swan-Tate cohomology $t_\G E$ (of a complex-oriented equivariant $E_\infty$ ring-spectrum $E$) is defined for any compact Lie group $G$, but when $G$ is $\T$, as below, things simplify considerably, for $t_\T H$ is then the periodic cyclic homology of the singular cochain algebra \cite{1, 7, 11}.  This account is concerned mostly with the structure of (the coefficient ring of) $t_\T H\Z$, but because $MU = E$ is the (torsion-free) universal example, it is convenient to start there (or skip to \S 2.3).  

{\bf 2.1} If $[X]$ is a compact cx-oriented $2d$-dimensional smooth manifold with circle action, perhaps with boundary, such that the restriction of the circle action to the boundary is free, \ie making the boundary quotient a $2(d-1)$ manifold endowed with a canonical circle bundle, it seems reasonable to call $[X/\T]$ a manifold with meromorphic circle action 
Then
\[
t^*_\T MU^* \ni [X/\T] \mapsto \partial_\T [X/\T] : = [(\partial X)/\T \to B\T]  \in MU_{*-2}B\T {\;},
\]
defines the boundary operator in the diagram\begin{footnote}{A preliminary account of such Hopf algebra biextensions can be found in the appendix to \S 11 of the 1959/60 Seminaire Cartan-Moore published in an adjacent timeline, but see more recently \cite{4}}\end{footnote}
\[
\xymatrix{ 
0 \ar[r] & E^*B\T \ar[d] \ar[r] & t^*_\T E \ar[r]^-{\partial_\T} \ar[d] & E_{*-2} B\T \ar[r] \ar[d] & 0 \\
0 \ar[r] & E^*[[c]] \ar[r] & E^*((c)) \ar[r]^b & E_*[b_*]  \ar[r] & 0}
\]
where $b \in {\rm Hom}_{\rm FG}(F_E,\G_m)$, \ie satisfying
\[
b(T) = 1 +\Sigma_{k \geq 1} b_k T^k {\;},{\;} b(T_0 +_E T_1) = b(T_0) \cdot b(T_1)  ,
\]
is the Ravenel-Wilson-Katz Cartier generating character for the formal group $F_E$ defined by the cx orientation of $E$. 

{\bf 2.2} The product of two such manifolds, \ie with circle actions free along the boundary, is a manifold with $\T \times \T$ structure; restricting along the diagonal $\T \to \T \times \T$ defines a manifold $[(X \times Y)/\T]$ with circle action, again 
free along the boundary
\[
\partial (X_0 \times X_1) = X_0 \times \partial X_1 \cup_{\partial X_0 \times \partial X_1} \partial X_0 \times X_1 {\;},
\]
defining a ring structure on $t^*_\T MU$. One wonders about Massey products.

{\bf 2.3} {\bf Examples} $E = H\Z {\;}\& {\;}K$

If $G$ is a finite group, $t^*_G K \cong R_\C(G) \otimes \Q$, whereas $t^*_\T K \cong (1-q)^{-1}K_\T$; where $K_\T \cong R_\C(\T)$ is Atiyah - Segal $\T$-equivariant $K$-theory (of a point), and $q :\T \subset \C^\times$ sends $q \to e^{i\theta}$. 
\bigskip

Recall from \cite{3}(Prop 3.4) that localization kills ($c$)-torsion, so
\[
t_\T H^*[M/\T] \cong H\Z^*({\rm Fix}_\T M) \otimes t_\T H\Z
\]
\ie with the grading scrambled by weights coming from normal bundles \cite{3}(\S 7); see \cite{5} for $t_{\Z_2} H\F_2$. Re $\SU(2)$, recall $B\SU(2) \cong {\mathbb H} P^\infty$ is not an $H$-space. 

{\bf 2.4} When $E = H\Z$ is Eilenberg - Mac Lane cohomology, we now have an exact sequence 
\[
\xymatrix{
0 \ar[r] & H^*B\T = \Z[c] \ar[r] & t_\T H\Z = \Z[c^{\pm 1}] \ar[r]^{\partial_\T} & H_{*-2} B\T = \Gamma[b_*] \ar[r] & 0 \\
0 \ar[r] &  c^3 \ar[r] & c^3  \ar[r] & 0 \\                                                                           
0 \ar[r] &  c^2 \ar[r] & c^2  \ar[r] & 0 \\
0 \ar[r] &  c \ar[r] & c \ar[r] & 0 \\
0 \ar[r] &  1 \ar[r] \ar[drr]^{\cong T \partial_\T = \partial^0_\T} & 1  \ar[r] & 0 \\
0 \ar[r] &  0 \ar[r] & c^{-1}  \ar[r] & 1 \\
0 \ar[r] &  0 \ar[r] & c^{-2}  \ar[r] & b_1 \\ 
0 \ar[r] &  0 \ar[r] & c^{-3} \ar[r] & b_2 \dots }
\] 

in which $\partial^0_\T : (\{0\} \in \C) \to (\C - \{0\}) = \C^\times \in {\rm free \;} \T$-spaces. The boundary operator thus defines a Rota-Baxter structure: invertible in negative degree and zero in positive degree.\bigskip

Here $\Gamma_*[b] = \Z[b_i \;|\; b_i b_k = (i,k) b_{I + k}]$ is the free divided power algebra on one generator; thus $b_k \mapsto b^k/k! \in \Gamma_*(b) \otimes \Q \cong \Q[b]$. The grading forces this sequence to split, as
\[
t^*_\T H = \Z[c] \oplus c^{-1}\Z[c^{-1}] 
\]
with $\partial_\T c^{-k} = b_{k-1}$, defining a representation of a Heisenberg group \cite{8}(\S 3).\bigskip

It is useful to think of $c^{-1}$ as the two-disk $D^2$ with its usual circle action; then 
\[
\partial_\T c^0 = 0, \partial_\T c^{-1} = 1, \partial_\T c^{-2} = [S^3/\T \cong S^2 \to \C P^\infty \cong B\T] = b \in MU_2 B\T {\;} ,
\]
making $[D^2/\T]$ to $[S^2/\T]$ much as open are to closed strings.\newpage

{\bf \S 3 Some applications}\bigskip

{\bf 3.1} The degrading functor $V_* \mapsto  V_* \otimes \Z((T))|_0  := V||T|| \in \Z((T)) -\Mod$ is defined on $\Z$ - graded modules by 
\[
\Hom^d(V,W) \ni \phi  \mapsto [\Sigma_{i > -\infty} v_i T^i  \to T^d\Sigma_{i > -\infty} \phi(v_i) T^i], {\;} v_i \in V_i
\]
so we have from \cite{5}(\S I) that
\[
b = - T^{-1} \log (1 - c^{-1}T) {\;} , {\;} c = - b^{-1}B^{-}(-bT) \in (t_\T H\Q)||T||
\]
with Bernoulli operator $B^{-}(D) = \frac{D}{e^D - 1}$, because
\[
(1 - c^{-1}T)^{-1} = \exp(bT) 
\]
(and more generally in the Lazard ring.) Note that $B_k/k = h_k$ are the cumulants of the uniform distribution.

These relations interpret $b$ as a de Rham current corresponding to the Noether charge $c$, \ie as something like its supporting cycle; but they make sense only after rationalizing. In particular, $b$ and $c$ have opposed natural variances. 
In this formalism $T$ is a kind of place-holder for a scaling action of Adams' operations in ${\mathbb N}^\times$, interpreted as an asymptotic expansion \cite{10}.
\bigskip

{\bf 3.2} The topological Mellin transform identifies $\Spec t_\T H\Z$ as a localization $\G_m = \Spec \Z[c^{\pm 1}]$ of the ungraded cohomology $H^*B\T \cong H^*\C P^\infty$:\bigskip
 
{\bf Proposition} {\it The differential form \cite{9}(\S 2.1 Proposition)
\[
d(bT) = - d \log( 1 - c^{-1}T) \in \Omega^1_\Z (t_\T H\Z)
\]
maps by $c^{-1}T \to e^\ve$ to the tempered distribution-valued one-form}
\[
d(bT) = \frac{d(c^{-1}T)}{1 - c^{-1}T} = \frac{e^\ve d\ve}{1 - e^\ve}  =  \li_0(\ve) \cdot d \ve = d \li_1(\ve) \in {\mathcal S}'(\R) \cdot d\ve \cong \Omega^1_{\rm temp}(\R) 
\]
where
\[
\li_0(x) = - (1 - e^{-x})^{-1} {\;} , {\;} \li_1(x) = - \log|1 - e^x| {\;}.
\]

{\bf 3.3} We interpret $db$ as Haar measure, translation-invariant with respect to $+_{\G_m}$, pulled back along the exponential map to the Lie algebra of $\G_m$, defining a moment map $b: \G_m \to \La$ (as in \cite{3}(\S 7)) for the circle. We can regard $b$ as defining a transcendental extension of $\C(c)$.

The resulting linear operator \cite{14}
\[
\pi^{\G_m}_! : \Omega^1_\Z (t_\T H\Z) \to  {\mathcal S}'(\R) \cdot d\ve 
\]
sends one-forms to de Rham Harvey-Lawson one-currents on the $\ve$-line, defining a pairing 
\[
{\mathcal S}(\R) \times \Z[c^{\pm 1}] \cdot dc \to \R \cdot d\ve
\]
in which $\ve = - \log c {\;} , {\;} d\ve = - c^{-1} dc$, suggesting $\ve$ as a measure of topological entropy or information loss. \bigskip

{\bf 4 Afterthoughts} \bigskip

{\bf 4.1} Convolutions of tempered distributions are not necessarily tempered, but the holomorphic family 
\[
\gamma^s_+ := \frac{x_+^{s-1}}{\Gamma(s)} \in {\mathcal S}'(\R)\{s\}
\]
of Gel'fand-\^Silov tempered distributions ($x \in \R$) represents the group of fractional differentiation operations $\partial^{-s}$ on functions such as $B^{-}(x) \sim (e^x - 1)^{-1} = \li_0(x)$.

Prop 3 of \cite{6} asserts that the function
\[
\li_s(x) = \gamma^s_+ \; * \li_0 = {\rm Li}_s(e^x) = 1 + \Sigma_{n \geq 1} \frac{e^{nx}}{n^s} \in C^\infty(\R_{<0})\{s\}, {\;} \li_s|_{x = 0} \mapsto \zeta(s)
\]
(as defined on the negative real line) extends to a holomorphic family $s \mapsto \li_s  \equiv - \pi \cot \pi s\cdot \gamma^s_+\in {\mathcal S}'(\R)\{s\}$ of tempered distributions on the whole real line. Consequently \cite{9}(\S 2.2)

{\it The divided moments
\[
\gamma^s_+ (\ve) * db = (\gamma^s_+  * d\li_1)(\ve)
\]
are tempered distributions. } \bigskip 

{\bf 4.2} Note that the distribution-valued `measure'  $\gamma^1_+ (\ve) * db(\ve)$ on $\R$ has no `value' at $x=0$, though the Taylor series of $\li_s (x)$ at $s = 1$ is
\[
(1 - s)^{-1} + \log |\frac{x}{e^x - 1}|(1 - s)^0  + \Sigma_{k \geq 1} t_k  \cdot (1-s)^k \dots
\]
with tempered distributions $t_*$. 

Nevertheless, these generalized moments may be useful in statistical mechanics. The Stefan/Boltzmann asymptotic expansion \cite{9}(\S 2.4),\cite{12}
\[
T^{-4} \boldsymbol{\rho} \sim 12 h\kappa^4 \cdot \gamma^4_+ * d\li_1
\]
(for the radiance\begin{footnote}{$\kappa = kh^{-1} \sim 20.836 \times 10^9$ Hertz per Kelvin}\end{footnote} $\boldsymbol{\rho}$ of an incandescent body in terms of its temperature) is how we understand the colors of the stars.

\newpage

\bibliographystyle{amsplain}

\end{document}